\documentclass[12pt,preprint]{emulateapj}
\begin{document}
\title{Low-luminosity galaxies in the early universe have
  observed sizes similar to star cluster complexes}
\author{R.J. Bouwens\altaffilmark{1}, G.D. Illingworth\altaffilmark{2}, P.G. van Dokkum\altaffilmark{3}, B. Ribeiro\altaffilmark{1}, P.A. Oesch\altaffilmark{4}, M. Stefanon\altaffilmark{1}}
\altaffiltext{1}{Leiden Observatory,
  Leiden University, NL-2300 RA Leiden, Netherlands}
\altaffiltext{2}{UCO/Lick Observatory, University of California, Santa
  Cruz, CA 95064}
\altaffiltext{3}{Department of Astronomy, Yale University, New Haven,
  CT 06520}
\altaffiltext{4}{Observatoire de Gen{\`e}ve, 1290 Versoix, Switzerland}
\begin{abstract}
We compare the sizes and luminosities of faint $z=6$-8 galaxies
magnified by the Hubble Frontier Fields (HFF) clusters with
star-forming regions, as well as more evolved objects, in the nearby
universe.  Our high-redshift comparison sample includes 333 $z=6$-8
galaxies, for which size measurements were made as part of a companion
study where lensing magnifications were estimated from various public
models.  Accurate size measurements for these sources are complicated
by the lens model uncertainties, but other results and arguments
suggest that faint galaxies are small, as discussed in a companion
study.  The measured sizes for sources in our comparison sample range
from $<$50 pc to $\sim$500 pc.  For many of the lowest luminosity
sources, extremely small sizes are inferred, reaching individual sizes
as small as 10-30 pc, with several sources in the 10-15 pc range with
our conservative magnification limits.  The sizes and luminosities are
similar to those of single star cluster complexes like 30 Doradus in
the lower-redshift universe and -- in a few cases -- super star
clusters.  The identification of these compact, faint star-forming
sources in the $z\sim6$-8 universe also allows us to set upper limits
on the proto-globular cluster LF at $z\sim6$.  By comparisons of the
counts and sizes with recent models, we rule out (with some caveats)
proto-globular cluster formation scenarios favoring substantial
($\xi=10$) post-formation mass loss and set useful upper limits on
others.  Our size results suggest we may be very close to discovering
a bona-fide population of forming globular clusters at high redshift.
\end{abstract}

\section{Introduction}

There are a wide variety of evolved stellar systems in the nearby
universe (Norris et al.\ 2014), from globular clusters (Brodie \&
Strader 2006; Kruijssen 2014; Renzini et al.\ 2015) to compact
elliptical galaxies (e.g., Faber 1973), ultra-faint dwarfs (e.g.,
Simon \& Geha 2007), and ultra-diffuse spheroids (e.g., van Dokkum et
al.\ 2015), each of which presumably has its own characteristic
formation pathway.  The high stellar densities in many of these
systems in combination with their old ages (e.g., Forbes \& Bridges
2010) suggest that the majority of their star formation occured at
$z\gtrsim 1.5$ when the gas densities in the universe were in general
much higher.

One potentially promising way forward for investigating the formation
of these local systems is by obtaining a sensitive, high-resolution
view of the distant universe.  Fortunately, such observations can be
obtained by combining the power of long exposures with the Hubble
Space Telescope with the magnifying effect of gravitational lensing,
as recently implemented in the ambitious Hubble Frontier Fields (HFF)
program (Coe et al.\ 2015; Lotz et al.\ 2017).  Such sensitive
observations allow us to probe to very low luminosities, as is likely
necessary to detect many of the progenitors of local systems.  The
high lensing magnifications from massive galaxy clusters stretch many
galaxies by substantial factors, allowing them to be studied at very
high spatial resolution.  As we discussed in Bouwens et al.\ (2021a),
this stretching can reliably be estimated up to linear magnifications
of $\sim$30$\times$ (or total magnification factors of
$\sim$50$\times$: see also Bouwens et al.\ 2017b where similar, though
smaller limits were presented with the then-current models, and
Meneghetti et al.\ 2017).

The promise of the HFF program to examine faint sources in the distant
universe at extremely high resolution was bourne out by the studies of
Kawamata et al.\ (2015) and Bouwens et al.\ (2017a), who identified
small $<$40 pc star-forming sources in the $z\sim6$-9 universe and
remarked that the size of such systems approached that of giant
molecular clouds or star-forming clumps in the local universe
(Kennicutt et al.\ 2003).  Vanzella et al.\ (2017a) speculated that a
few of these small star-forming sources behind the HFF clusters could
correspond to proto-globular clusters.  They supported their
hypothesis with MUSE spectroscopy of the sources, noting small
probable dynamical masses (due to the small measured velocity
dispersions) and probable physical associations with brighter
neighbors (due to their similar redshifts).  In two follow-up studies,
Vanzella et al.\ (2019, 2020) identified two candidate proto-globular
cluster systems with inferred sizes $<$13 pc and 5-10 pc, which is
very close to the expected sizes of young massive clusters (Kruijssen
2014).

We can now build on this recent work, using large catalogs of
$z\sim6$-9 galaxies constructed from the deep HST observations over
the six HFF clusters.  Lensing models have also improved, as a result
of a larger number of approaches and improved techniques.  The median
of these models is expected to be more robust than any individual
model and the scatter provides an estimate of the uncertainties.
Thanks to the improvements in the lensing models and the large number
of clusters targeted, it has become possible to measure the sizes for
large samples of faint $z=6$-9 galaxies and to very small sizes.
Kawamata et al.\ (2018) report results for 181 $z=6$-9 galaxies, while
Bouwens et al.\ (2021a) report results for 333 $z=6$-8 galaxies.  Both
studies report results on all six HFF clusters and find substantial
numbers ($>$30) of $z=6$-9 star-forming sources with plausible
half-light radii $\leq$50 pc.  This remains the case even when we
adopt the conservative approach discussed in Bouwens et al.\ (2021a) to
limit linear and total magnification factors to 30 and 50,
respectively.

Given the small inferred sizes of the fainter lensed sources
identified by Kawamata et al.\ (2018) and Bouwens et al.\ (2021a), it
is interesting to place these sources in the context of various
stellar systems that they may evolve into today, as well as other
small star-forming systems like star clusters or cluster complexes.
An initial look at such comparisons were already executed in an
earlier unpublished study by our group (Bouwens et al.\ 2017) and also
by Kikuchihara et al.\ (2020).  An important early inference from
these studies was that lensed $z=6$-8 galaxies have sizes and masses
that appear to lie in the range of $\sim$50-500 pc and 10$^7$ to
10$^8$ $M_{\odot}$, lying somewhere between ultra compact
dwarfs/globular clusters and compact elliptical galaxies in size/mass
space.

The purpose of the present work is to revisit these comparisons,
utilizing the large selection of $z=6$-8 galaxies with size
measurements from Bouwens et al.\ (2021a) while leveraging a variety of
new results -- both from simulations (Ploeckinger et al.\ 2019;
Pfeffer et al.\ 2019) and from local observations (Zick et al.\ 2018)
-- to allow for an improved interpretation of the observational
results.  Two aspects are of particular interest: (1) comparing the
sizes of sources in these selections with star cluster complexes like
30 Doradus in the LMC and (2) using the Bouwens et al.\ (2021a)
searches to set constraints on proto-globular cluster formation models
such as described by Boylan-Kolchin (2017), Pfeffer et al.\ (2019),
J. Pfeffer et al.\ (2021, in prep), and Pozzetti et al.\ (2019).
Throughout the paper, we assume a standard ``concordance'' cosmology
with $H_0=70$ km s$^{-1}$ Mpc$^{-1}$, $\Omega_{\rm m}=0.3$ and
$\Omega_{\Lambda}=0.7$, which is in good agreement with recent
cosmological constraints (Planck Collaboration et al.\ 2016).
Magnitudes are in the AB system (Oke \& Gunn 1983).

\begin{deluxetable*}{ccccccc}
\tablecolumns{7}
\setlength{\tabcolsep}{14pt}
\tabletypesize{\footnotesize}
\tablewidth{17.8cm}
\tablecaption{Catalog of Tiny Star-Forming Sources\tablenotemark{a,b} at $z\sim6$-8\label{tab:tiny}}
\tablehead{\colhead{ID} & \colhead{R.A.} & \colhead{Decl} & \colhead{$M_{UV}$} & \colhead{$\mu$\tablenotemark{c}} & \colhead{$\mu_{1D}$\tablenotemark{d}} & \colhead{$r_e$ (pc)\tablenotemark{e}}}
\startdata
A2744Z-4222023578  &  00:14:22.20  &  $-$30:23:57.8  & $-$15.6$_{-1.2}^{+0.7}$  & 50.1$_{-33.6}^{+41.4}$  & 8.5$_{-3.8}^{+4.4}$  & 31$_{-13}^{+31}$\\
A2744Z-4222023578  &  00:14:22.20  &  $-$30:23:57.8  & $-$15.6  & 50.0\tablenotemark{f}  & 8.5$_{-3.8}^{+4.4}$  & 31$_{-13}^{+31}$\\
A2744Y-4204124034*  &  00:14:20.41  &  $-$30:24:03.4  & $-$14.0$_{-1.6}^{+1.2}$  & 75.0$_{-58.0}^{+151.2}$  & 15.1$_{-9.5}^{+24.0}$  & 14$_{-9}^{+32}$\\
A2744Y-4204124034*  &  00:14:20.41  &  $-$30:24:03.4  & $-$14.4  & 50.0\tablenotemark{f}  & 15.1$_{-9.5}^{+24.0}$  & 14$_{-9}^{+32}$\\
M0416I-6095704260*  &  04:16:09.57  &  $-$24:04:26.0  & $-$15.0$_{-0.2}^{+0.3}$  & 12.5$_{-2.5}^{+3.3}$  & 8.0$_{-2.0}^{+2.5}$  & 28$_{-9}^{+13}$\\
M0416I-6118103480\tablenotemark{$\ddagger$}  &  04:16:11.81  &  $-$24:03:48.1  & $-$15.0$_{-1.0}^{+1.3}$  & 33.6$_{-19.9}^{+81.6}$  & 24.5$_{-15.2}^{+85.1}$  & 16$_{-13}^{+43}$\\
M0416I-6130803432*\tablenotemark{$\ddagger$}  &  04:16:13.08  &  $-$24:03:43.2  & $-$17.3$_{-0.2}^{+0.2}$  & 3.5$_{-0.6}^{+0.7}$  & 3.4$_{-0.6}^{+0.3}$  & 38$_{-4}^{+8}$\\
M0416I-6115434445\tablenotemark{$\dagger$,$\ddagger$}  &  04:16:11.54  &  $-$24:03:44.5  & $-$14.5$_{-1.0}^{+0.9}$  & 43.0$_{-25.4}^{+54.2}$  & 31.0$_{-18.4}^{+48.0}$  & 21$_{-14}^{+44}$\\
M0416I-6115434445\tablenotemark{$\dagger$,$\ddagger$}  &  04:16:11.54  &  $-$24:03:44.5  & $-$14.5$_{-1.0}^{+0.9}$  & 43.0$_{-25.4}^{+54.2}$  & 30.0\tablenotemark{g}  & 22\\
M0416I-6106703335*  &  04:16:10.67  &  $-$24:03:33.5  & $-$16.3$_{-0.2}^{+0.3}$  & 11.1$_{-2.0}^{+3.7}$  & 8.8$_{-0.7}^{+3.1}$  & 33$_{-11}^{+10}$\\
M0416Z-6114803434  &  04:16:11.48  &  $-$24:03:43.5  & $-$17.0$_{-0.5}^{+0.3}$  & 21.3$_{-7.6}^{+6.3}$  & 14.9$_{-4.8}^{+7.8}$  & 38$_{-16}^{+26}$\\
M0416Y-6088104378*  &  04:16:08.81  &  $-$24:04:37.9  & $-$13.4$_{-1.1}^{+1.0}$  & 62.1$_{-39.5}^{+90.8}$  & 38.3$_{-27.0}^{+45.7}$  & 11$_{-7}^{+33}$\\
M0416Y-6088104378*  &  04:16:08.81  &  $-$24:04:37.9  & $-$13.7  & 50.0\tablenotemark{f}  & 30.0\tablenotemark{g}  & 14\\
M0717Z-7354743496  &  07:17:35.47  &  37:43:49.7  & $-$13.9$_{-1.4}^{+1.6}$  & 57.7$_{-42.1}^{+205.4}$  & 46.8$_{-34.0}^{+142.4}$  & 27$_{-21}^{+103}$\\
M0717Z-7354743496  &  07:17:35.47  &  37:43:49.7  & $-$14.0  & 50.0\tablenotemark{f}  & 30.0\tablenotemark{g}  & 42\\
M0717I-7374244282  &  07:17:37.42  &  37:44:28.2  & $-$15.0$_{-0.8}^{+0.8}$  & 9.6$_{-5.0}^{+10.8}$  & 3.6$_{-0.9}^{+1.5}$  & 34$_{-12}^{+18}$\\
M0717Z-7390844017\tablenotemark{$\ddagger$}  &  07:17:39.08  &  37:44:01.7  & $-$17.1$_{-1.8}^{+0.4}$  & 21.8$_{-17.8}^{+10.7}$  & 17.7$_{-13.3}^{+14.4}$  & 24$_{-12}^{+85}$\\
M0717Z-7401344384  &  07:17:40.13  &  37:44:38.5  & $-$16.1$_{-0.7}^{+0.0}$  & 8.3$_{-3.9}^{+0.3}$  & 6.4$_{-2.3}^{+1.5}$  & 39$_{-9}^{+24}$\\
M0717Z-7311744437*  &  07:17:31.17  &  37:44:43.8  & $-$16.0$_{-0.4}^{+0.8}$  & 8.2$_{-2.6}^{+8.8}$  & 4.0$_{-0.7}^{+3.1}$  & 27$_{-14}^{+15}$\\
M0717Y-7336744331  &  07:17:33.67  &  37:44:33.1  & $-$14.4$_{-1.3}^{+2.3}$  & 43.1$_{-29.9}^{+312.6}$  & 7.6$_{-3.7}^{+40.9}$  & 31$_{-26}^{+58}$\\
M0717Y-7329744137  &  07:17:32.97  &  37:44:13.7  & $-$15.3$_{-0.6}^{+0.9}$  & 13.7$_{-5.9}^{+16.8}$  & 7.5$_{-3.7}^{+8.6}$  & 34$_{-20}^{+48}$\\
M1149I-9384023344  &  11:49:38.40  &  22:23:34.5  & $-$13.0$_{-2.4}^{+2.6}$  & 124.7$_{-111.1}^{+1226.2}$  & 90.7$_{-81.8}^{+949.9}$  & 26$_{-24}^{+342}$\\
M1149I-9384023344  &  11:49:38.40  &  22:23:34.5  & $-$14.0  & 50.0\tablenotemark{f}  & 30.0\tablenotemark{g}  & 79\\
M1149Y-9377423253*  &  11:49:37.74  &  22:23:25.4  & $-$14.9$_{-0.3}^{+0.1}$  & 16.2$_{-4.1}^{+2.0}$  & 7.9$_{-3.9}^{+1.3}$  & 19$_{-4}^{+20}$\\
A370I-9544334397*  &  02:39:54.43  &  $-$1:34:39.7  & $-$12.7$_{-1.2}^{+3.5}$  & 228.9$_{-151.9}^{+5472.8}$  & 23.1$_{-11.0}^{+484.7}$  & 10$_{-10}^{+21}$\\
A370I-9544334397*  &  02:39:54.43  &  $-$1:34:39.7  & $-$14.4  & 50.0\tablenotemark{f} & 23.1$_{-11.0}^{+484.7}$  & 10$_{-10}^{+21}$\\
A370Z-9535934395  &  02:39:53.59  &  $-$1:34:39.6  & $-$15.5$_{-0.1}^{+2.1}$  & 10.9$_{-1.2}^{+65.5}$  & 7.7$_{-0.8}^{+35.5}$  & 31$_{-26}^{+28}$\\
AS1063I-8439532143  &  22:48:43.95  &  $-$44:32:14.3  & $-$13.5$_{-2.8}^{+0.6}$  & 42.3$_{-39.0}^{+28.4}$  & 19.0$_{-16.3}^{+12.6}$  & 33$_{-15}^{+219}$\\
AS1063I-8469531055  &  22:48:46.95  &  $-$44:31:05.5  & $-$14.9$_{-1.8}^{+0.2}$  & 12.9$_{-10.4}^{+2.3}$  & 10.3$_{-7.9}^{+1.8}$  & 35$_{-7}^{+115}$\\
AS1063Y-8436531406*  &  22:48:43.65  &  $-$44:31:40.6  & $-$12.5$_{-1.6}^{+1.0}$  & 175.3$_{-134.0}^{+257.9}$  & 77.7$_{-57.8}^{+81.7}$  & 9$_{-5}^{+30}$\\
AS1063Y-8436531406*  &  22:48:43.65  &  $-$44:31:40.6  & $-$13.9  & 50.0\tablenotemark{f}  & 30.0\tablenotemark{g}  & 23
\enddata
\tablenotetext{a}{All sources with inferred half-light radii less than 40 pc are included in this table.  As we discuss in \S3 and \S4, these small star-forming sources could be characterized as star cluster complexes, super star clusters, proto-globular clusters, or especially compact galaxies.}
\tablenotetext{b}{In cases where the median total magnification and
  linear magnification exceeds 50 and 30, resepectively, we quote
  alternative estimates for the absolute magnitude $M_{UV}$ and
  physical size $r_{e}$ with the total and linear magnification fixed
  to 50 and 30, respectively.  This alternate estimates of $M_{UV}$
  and $r_e$ are provided for sources given the challenges in relying
  upon magnification factors in excess of these values (\S4.2 from
  Bouwens et al.\ 2021a and Bouwens et al.\ 2017b).}
\tablenotetext{c}{Median magnification factors (and $1\sigma$ uncertainties) derived weighting equally the latest public version 3/4 parametric models from each lensing methodology (\S4.1 from Bouwens et al.\ 2021a).}
\tablenotetext{d}{$\mu_{1D}$ are the median one-dimensional magnification factors (and $1\sigma$ uncertainties) along the major shear axis $\mu^{1/2} S^{1/2}$ weighting equally the parametric models from each lensing methodology.  This is the same quantity as $\mu_{tang}$ reported by Vanzella et al.\ (2017a).}
\tablenotetext{e}{Inferred half-light radius in physical units.  The quoted uncertainties include both uncertainties in the spatial fits and uncertainties in the lensing models.  These half-light radii are derived by fitting a Sersic profile to the imaging observations while fixing the Sersic parameter $n$ to 1, as discussed in Bouwens et al.\ (2021a).}
\tablenotetext{f}{Total magnification factor $\mu$ set to 50.0 where the median magnification models begin losing their predictive power (Bouwens et al.\ 2021a).  Because of this use of a limit on the magnification factor, no uncertainty is quoted on the inferred $UV$ luminosity $M_{UV}$.  Nevertheless, a useful indication of the uncertainties can be seen, if we take the median $1\sigma$ uncertainties in the quoted magnification factors before imposing a limit, which would give a value of 0.59 dex.}
\tablenotetext{g}{Linear magnification factor $\mu_{1D}$ set to 30.0 where the median magnification models begin losing their predictive power (Bouwens et al.\ 2021a).  Because of this use of a limit on the magnification factor, no uncertainty is quoted on the inferred size $r_e$.  Nevertheless, a useful indication of the uncertainties can be seen, if we take the median $1\sigma$ uncertainties in the quoted magnification factors before imposing a limit, which would give a value of 0.52 dex.}
\tablenotetext{*}{One of the sources where the upper $1\sigma$ limit on the inferred half-light radius is less than 50 pc.}
\tablenotetext{$\dagger$}{Tiny star-forming source also presented in Vanzella et al.\ (2017a).}
\tablenotetext{$\ddagger$}{Source also has an inferred size of $\leq$40 pc in  the Kawamata et al.\ (2018) catalog.}
\end{deluxetable*}

\section{High-Redshift Samples and Size Measurements}

Here we provide a brief summary of the $z=6$-8 selections from Bouwens
et al.\ (2021a) that we utilize in interpreting their properties
relative to various star-forming or evolved sources in the local
universe.  For a more detailed description of the procedure used in
selecting sources or deriving sizes/luminosities, the interested
reader is referred to our companion paper (Bouwens et al.\ 2021a).

\subsection{Selection and Size Measurements for $z\sim6$-8 Galaxies}

Briefly, the $z\sim6$-8 samples we utilize are selected from the v1.0
reductions of the HST data over all six HFF clusters.  A combination
of Lyman-break and photometric-redshift selection criteria are applied
in creating source catalogs from the HST images.  Bright foreground
galaxies and also the intracluster light were modeled and subtracted
from the HST images, before construction of the catalogs.

In deriving the luminosities and sizes of individual sources, source
magnifications $\mu$ and shear factors $S$ are estimated from all of
the v3 or v4 public parametric lensing models.  The shear factor $S$
utilized here was defined as
\begin{equation}
S = \left\{
\begin{array}{lr}
\frac{1-\kappa - \gamma}{1-\kappa+\gamma}, &
\text{for } \frac{1-\kappa - \gamma}{1-\kappa+\gamma} \geq 1\\
\frac{1-\kappa + \gamma}{1-\kappa-\gamma}, &
\text{for } \frac{1-\kappa - \gamma}{1-\kappa+\gamma} < 1
\end{array}\right.
\end{equation}
(see Bouwens et al.\ 2017a and Bouwens et al.\ 2020).  $\kappa$ is the
convergence and $\gamma$ is the shear.  From the estimated
magnifications $\mu$ and shear factors $S$, we can derive a median
linear magnification along the major and minor shear axes, i.e., $(\mu
S)^{1/2}$ and $(\mu / S)^{1/2}$, which we use in measuring the
physical size of the source along both axes.  We find that these
linear magnification factors can now be reliably estimated to
magnification factors of $\sim$30 (\S4.2 from Bouwens et al.\ 2020)
and possibly to slightly larger factors using a median of the latest
parametric lensing models available for the HFF clusters (e.g., as in
Livermore et al.\ 2017; Bouwens et al.\ 2017b).

The fits to the sizes of individual sources are then performed using an
Markov-Chain Monte-Carlo procedure, where we take a model Sersic
profile, transform the profile according to the median magnification
factors, convolve the derived profile with the PSF, compare the result
with the observations, and compute a residual.  We use our MCMC
procedure to minimize the sum of the square of the residuals.

A full compilation of the coordinates, luminosities, and sizes of the
$z\sim6$-8 galaxies Bouwens et al.\ (2021a) identify behind all six
HFF clusters is given in their Table 2.  The uncertainties Bouwens et
al.\ (2021a) compute on the derived sizes include both the formal
uncertainties on the size fits and the $1\sigma$ error $\mu_{1D}$
computed based on the range in linear magnifications predicted by the
parametric lensing models.  Table~\ref{tab:tiny} of Bouwens et
al.\ (2021a) tabulates the sizes and luminosities of those sources
with linear $\mu_{1D}$ and total magnification factors $\mu$ $>$30 and
$>$50, respectively, a second time but setting the magnification
factors to 30 and 50, respectively, to provide more conservative size
and luminosity estimates, given the breakdown in the predictive power
of these models at very high magnification factors.  When adopting an
upper limit of 30 and 50 on the linear and total magnification
factors, respectively, no uncertainties are quoted on the sizes
$r_{e}$ or luminosities $M_{UV}$, respectively, of sources.
Nevertheless, if we take the median $1\sigma$ scatter in the quoted
magnification factors before imposing these limits, the typical
uncertainty would be 0.52 and 0.59 dex, respectively.  The total
magnification $\mu$ and linear magnification $\mu_{1D}$ are treated as
independent for the purposes of imposing the upper limits.

There are 23 objects from the Bouwens et al.\ (2021a) compilation with
sizes $<$40 pc, with the three smallest sources having sizes of
11$_{-7}^{+33}$ pc (M0416Y-6088104378), 11$_{-7}^{+33}$ pc
10$_{-10}^{+21}$ (A370I-9544334397), and 9$_{-5}^{+30}$ pc
(AS1063Y-8436531406).  For convenience, we include a list of these
sources in Table~\ref{tab:tiny}.  The uncertainties in the measured
sizes for these small sources are clearly substantial.  Of the 23
sources in Table~\ref{tab:tiny}, only three have measured sizes less
than 40 pc after allowing for the positive $1\sigma$ fluctuations in
the measured sizes.  Nine sources are within an upper bound of 50 pc
(a slightly less stringent limit) allowing for the $1\sigma$
uncertainties.  These 9 sources are indicated in Table~\ref{tab:tiny}
with a ``*'' to indicate that their small size measurements are more
robust.  As in Table 2 of Bouwens et al. (2021a), we also include
sources a second time when their linear and total magnification
factors exceed 30 and 50, respectively, but setting these factors to
30 and 50, respectively.

\subsection{Comparison with Previous Compilations of Small Star-Forming Sources}

Here we compare the sizes of small galaxies in our $z\sim 6$-8 sample
to similarly small ($\sim$10-100 pc) sources identified by Vanzella et
al.\ (2017a) and to new results recently obtained by Kawamata et
al.\ (2018: which are an update to the earlier Kawamata et al.\ 2015
results).  In the companion paper to this one (Bouwens et al.\ 2021a),
we provided a similar comparison but focusing on our entire lensed
sample of $z\sim6$-8 sources.

Two of the three star-forming candidates that Vanzella et al.\ (2017a)
identify over MACS0416 are also included in our compilation of small
sources (Table~\ref{tab:tiny}).  M0416I-6115434445 corresponds to GC1
from Vanzella et al.\ (2017a), while M0416Z-6114803434 corresponds to
D1 from Vanzella et al.\ (2017a) and Vanzella et al.\ (2019).  D2 from
Vanzella et al.\ (2017a) is also in the Bouwens et al.\ (2021a) source
catalogs.

For GC1, D1, and D2, we measure half-light radii of 21$_{-14}^{+44}$
pc, 38$_{-16}^{+26}$ pc, 72$_{-4}^{+13}$ pc, while Vanzella et
al.\ (2017a) measure 16$\pm$7 pc, 140$\pm$13 pc, and $<$100 pc,
respectively, for these.  Vanzella et al.\ (2019) later revisited
their structural analysis of D1 and derived a half-light radius of
$<$13 pc for the core of D1 and 44 pc for the source as a whole.  The
sizes we infer for M0416I-6115434445/GC1 and M0416I-6103003258/D2 are
in excellent agreement with those from Vanzella et al.\ (2017a).  For
the third source M0416Z-6114803434/D1, the half-light radius Bouwens
et al.\ (2021a) estimate is in good agreement with the size Vanzella et
al.\ (2019) derive for the source as a whole.

The larger size estimate for M0416Z-6114803434/D1 from Vanzella et
al.\ (2017a) likely results from the $n=3.0\pm0.3$ Sersic parameter
they find from their size fits, which might suggest a larger size for
the source.  Vanzella et al.\ (2019), however, in a follow-up study
argue that perhaps a preferable interpretation of observations is in
terms of a two component fit, with a bright central component for the
source.  The presence of a compact, central high surface-brightness
region with an estimated size of $<$13 pc is fascinating in the
context of our discussion in \S\ref{sec:starcluster} and its
relevance to the suspected presence of high surface brightness
star-forming complexes as per Figure~\ref{fig:cartoon}.  This
conceptual model is discussed in more depth in
\S\ref{sec:starcluster}.

We can also compare with the compilation of small $z=6$-8 sources
identified by Kawamata et al.\ (2018).  Of the 23 $z=6$-8 sources in
the Kawamata et al.\ (2018) catalogs with source sizes less than 50 pc
or smaller and which also appear in our catalogs, we measure a median
size of 63 pc.  For the eight $z=6$-8 galaxies in our catalogs with
inferred source sizes $<$50 pc which also appear in the Kawamata et
al.\ (2018) catalogs, the median size Kawamata et al.\ (2018) derive
is 35 pc.  This demonstrates that the derived sizes for the smallest
sources in each of our catalogs is reproducible (at least broadly
speaking).  Overall, for the 24 sources in our collective $z=6$-8
catalogs where Kawamata et al.\ (2018) or we derive sizes less than 50
pc, we compute an RMS source-to-source scatter of 0.31 dex.  This is
similar to the 0.30-dex scatter we find in size measurements for the
92 sources in common between our samples.  The size comparisons with
Kawamata et al.\ (2018) are quite reassuring.

We remark that the higher significance $5\sigma$ detections that
Kawamata et al.\ (2018) require in multiple near-IR bands for their
own selections of $z=6$-9 galaxies appear to be a major reason why
their catalogs include a smaller number of sources (181 sources) than
our own (333 sources).  In our own selections, we only require sources
be detected at 6$\sigma$, based on the $\chi^2$ parameter results we
obtain by adding in quadrature the detection significance in the
$Y_{105}J_{125}JH_{140}H_{160}$ near-IR bands.  We made use of a
similar procedure in a variety of previous analyses (e.g., Bouwens et
al.\ 2011, 2015, 2021b).  This results in our selection including many
more sources at $>$28 mag than are present in the Kawamata et
al.\ (2018) selection and thus being $\sim$1.8$\times$ larger.

\begin{figure*}
\epsscale{1.11}
\plotone{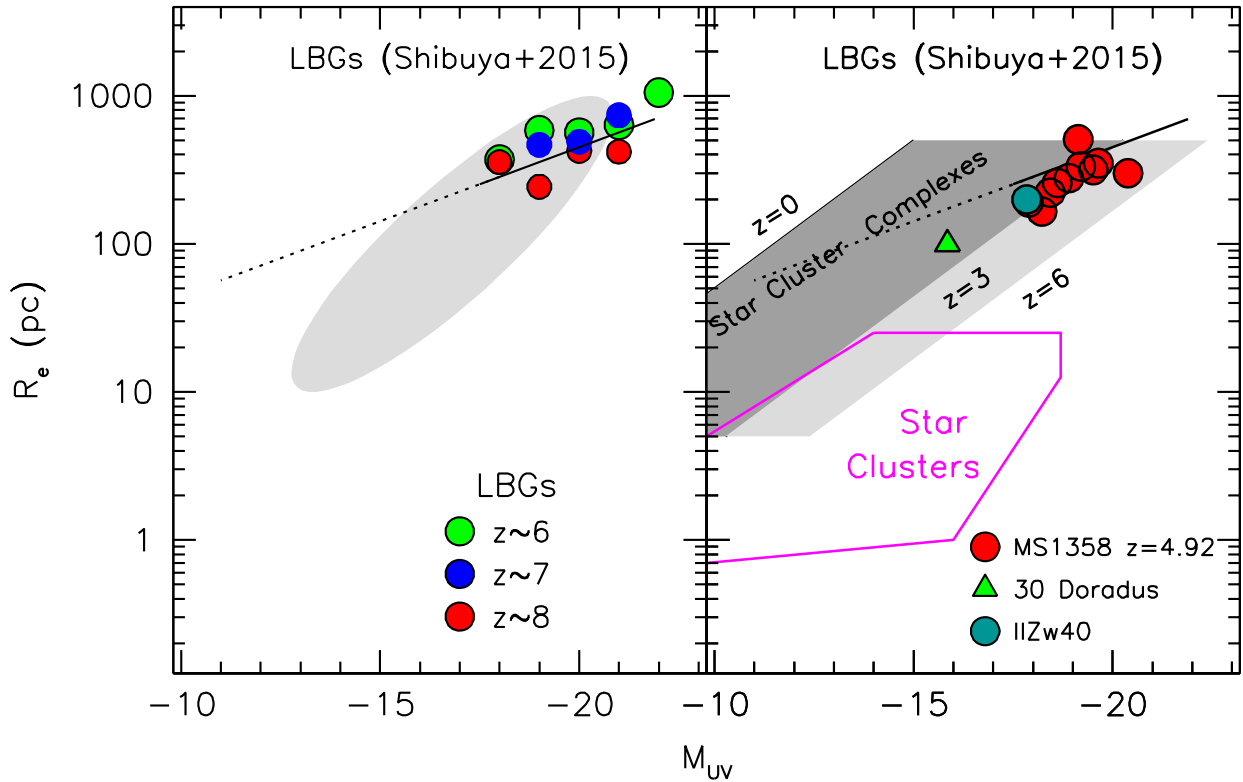}
\caption{Median size vs. luminosity relation of galaxies identified in
  blank field studies, i.e., the XDF/HUDF and CANDELS (\textit{left
    panel}) and those of star cluster complexes (\textit{right
    panel}).  The canonical size-luminosity relation for high-redshift
  galaxies is presented using both the Shibuya et al.\ (2015) fit
  results (\textit{black line}) and median sizes at $z\sim6$
  (\textit{green circles}), $z\sim7$ (\textit{blue circles}), and
  $z\sim8$ (\textit{red circles}).  The black dotted line shows an
  extrapolation of the best-fit Shibuya et al.\ (2015) $r\propto
  L^{0.26\pm0.03}$ trend to lower luminosities, as Bouwens et
  al.\ (2021a) find in their \S3.  The shaded grey area shows the
  approximate region occupied by lensed $z\sim6$-8 sources identified
  by Bouwens et al.\ (2021a).  The right panel shows the regions in
  size-luminosity space occupied by star clusters (and super star
  clusters) and cluster complexes.  The dark gray region indicates the
  size-luminosity relation for star cluster complexes in $z=0$-3
  galaxies inferred by Livermore et al.\ (2015) by fitting to the
  $z=0$ results from SINGS (Kennicutt et al.\ 2003), as well as the
  results of Jones et al.\ (2010), Swinbank et al.\ (2012), Livermore
  et al.\ (2012), Wisnioski et al.\ (2012), and Livermore et
  al.\ (2015).  The light gray region indicates the size-luminosity
  relation for star cluster complexes extrapolating this relation to
  $z=3$-6.  The solid red circles correspond to the measured sizes and
  equivalent $UV$ luminosities of the star cluster complexes
  identified in the highly magnified $z=4.92$ galaxy behind MS1358+62
  (Franx et al.\ 1997) by Swinbank et al.\ (2009) and Jones et
  al.\ (2010), while the solid green triangle and cyan circle
  correspond to the sizes and luminosities of 30 Doradus and IIZw40,
  respectively (Meylan 1993; Vanzi et al.\ 2008).  The magenta lines
  enclose the luminosities and sizes measured for star clusters and
  super star clusters at $z\sim0$ (Meurer et
  al.\ 1995).\label{fig:msre00}}
\end{figure*}

\begin{figure*}
\epsscale{0.95}
\plotone{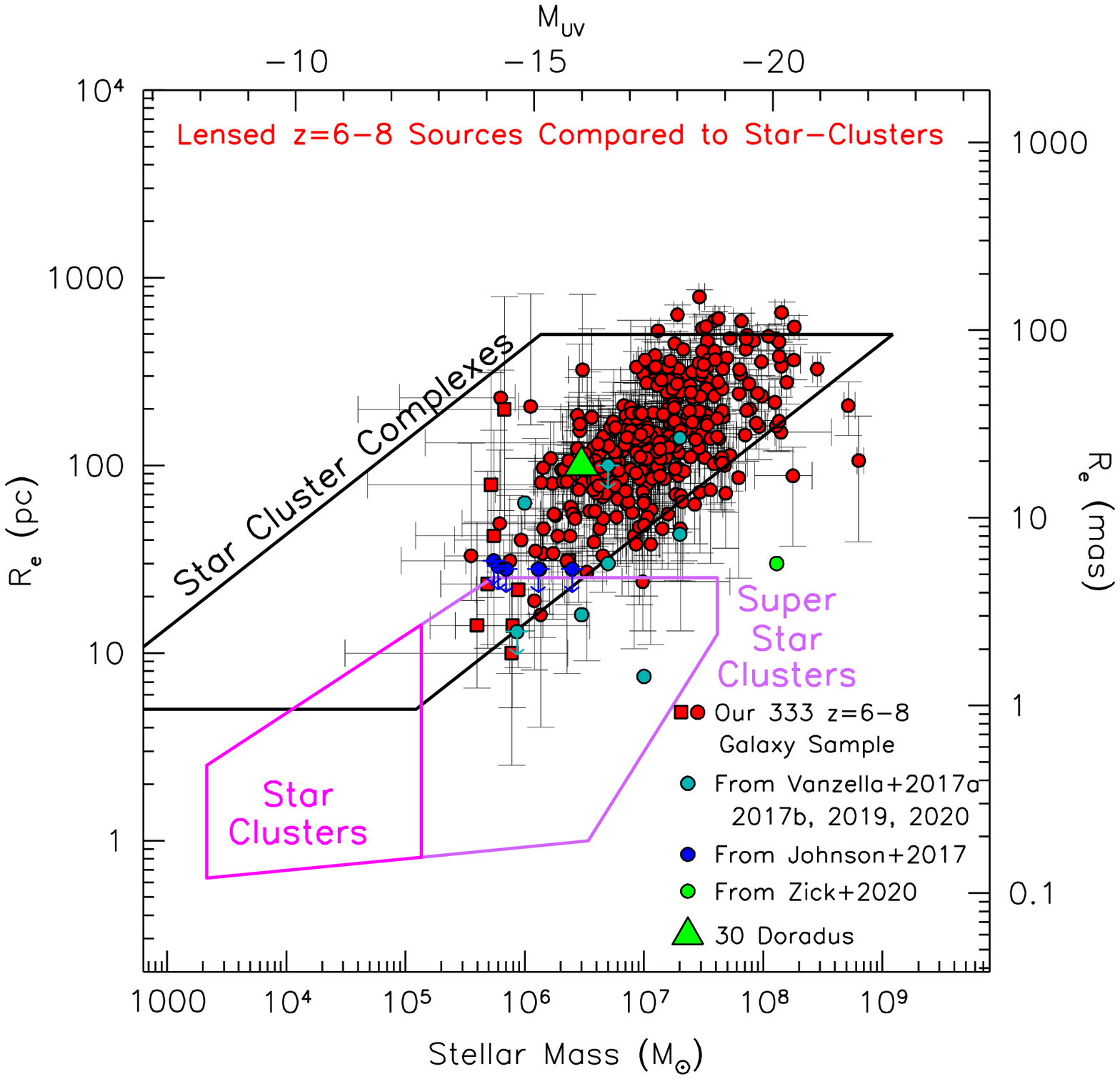}
\caption{Comparison of the inferred sizes [in pc] and luminosities of
  lensed galaxies in the HFF clusters (\textit{red circles}) with star
  clusters (\textit{demarcated by the magenta lines:}
  \S\ref{sec:intro_cluster}), super star clusters (\textit{demarcated
    by the violet lines:} \S\ref{sec:intro_starcluster}), and star
  cluster complexes (\textit{demarcated by the black lines:}
  \S\ref{sec:intro_clustercomplex} and Figure~\ref{fig:msre00}).  The
  right vertical axis shows the corresponding angular size of sources
  in mas if they are at $z\sim7$.  $1\sigma$ errors on the inferred
  sizes and $UV$ luminosities are quoted based on the 68\% confidence
  intervals in the size fits and the range of magnification factors in
  the parametric lensing models, as in Figure 4 (middle panel) of
  Bouwens et al.\ (2021a).  For those sources where the median total
  and linear magnification factors exceed 30 and 50, respectively, the
  sizes and luminosities are derived assuming maximum magnification
  factors of 30 and 50, respectively (but keeping the same
  uncertainties on the size and luminosity measurements as they were
  estimated prior to imposing upper limits on the magnification
  factors).  The sources where upper limits have been applied to the
  magnification factors are shown with the red squares.  The cyan,
  blue, and green circles show the sizes and luminosities reported by
  proto-globular clusters and star cluster candidates claimed by
  Vanzella et al.\ (2017a, 2017b, 2019, 2020), Johnson et al.\ (2017),
  and Zick et al.\ (2020) while the green triangle shows the size and
  luminosity of the 30 Doradus star complex.  The conversion between a
  given $UV$ luminosity and a stellar mass is made assuming a star
  formation duration of 10 Myr.  While most of the lensed $z=6$-8
  sources in the HFF observations appear to have sizes and
  luminosities consistent with star cluster complexes seen in $z=0$-3
  galaxies (Bastian et al.\ 2006; Jones et al.\ 2010; Wisnioski et
  al.\ 2012; Swinbank et al.\ 2012; Livermore et al.\ 2012; Adamo et
  al.\ 2013; Livermore et al.\ 2015; Johnson et al.\ 2017;
  Dessauges-Zavadsky et al.\ 2017), a few sources lie in the super
  star cluster region of parameter space.  We use the few sources
  found in the star cluster region to place constraints on the volume
  density of proto-globular clusters at $z\sim6$ (\S\ref{sec:gclf} and
  Figure~\ref{fig:gclf}).\label{fig:msre_sc}}
\end{figure*}

\section{Comparison to Star Clusters, Super Star Clusters, and Star Cluster Complexes\label{sec:starcluster}}

Given the extremely small sizes (i.e., $<$40 pc) inferred for some
fraction (here, 7\% of our sample of 333 galaxies) of the lensed
sources identified behind the HFF clusters (Bouwens et al.\ 2017a;
Vanzella et al.\ 2017a, 2017b, 2019), it is interesting to compare
these sources with more familiar star-forming entities in the nearby
universe in terms of their sizes and luminosities.  These entities
broadly fall into two classes by size, star clusters and star cluster
complexes.

\subsection{Star Clusters and Star Cluster Complexes\label{sec:intro_cluster}}

\subsubsection{Star Clusters and Super Star Clusters\label{sec:intro_starcluster}}

The first of these classes are the smaller star-forming objects, star
clusters and super star clusters.  Meurer et al.\ (1995) provide a
convenient summary of where star clusters lie in terms of their
effective radii and $UV$ luminosities $M_{UV}$ in their Figure 14.
$UV$ luminosities of the star clusters extend from $-$9 mag to $-19$
mag, masses range from $10^4$ to $10^8$ $M_{\odot}$ (Maraston et
al.\ 2004; Cabrera-Ziri et al.\ 2014, 2016), while the typical
effective radii of star clusters range from 0.5 pc to 4 pc (e.g., Lada
\& Lada 2003).

The magenta line in the right panel of Figure~\ref{fig:msre00}
demarcates the approximate sizes and luminosities found for star
clusters and super star clusters in the nearby universe.  For
comparison, the size-luminosity relation derived in Bouwens et
al.\ (2021a: \S3) for bright star-forming galaxies at $z=6$-8 is also
presented in the left panel of Figure~\ref{fig:msre00}, with an
extrapolation to lower luminosities.  We also show broadly, as a grey
shaded area, the region occupied by a fainter sample of 333 galaxies,
from Bouwens et al.\ (2021a).

The most massive ($>$10$^5$ $M_{\odot}$) star clusters are often
called super star clusters, with the effective radii extending to a
maximum size of $\sim$20 pc (e.g., Meurer et al.\ 1995; Rejkuba et
al.\ 2007; Murray 2009; Bastian et al.\ 2013).  Meurer et al.\ (1995)
classify any star clusters with $UV$ luminosities greater than $-$14
mag as super star clusters.  We refer the interested readers to other
excellent discussions on this topic by Rejkuba et al.\ (2007), Murray
(2009), and Bastian et al.\ (2013).

\subsubsection{Star Cluster Complexes\label{sec:intro_clustercomplex}}

The larger class of star-forming regions are star cluster complexes
seen in star-forming galaxies at $z=0$ (Kennicutt et al.\ 2003;
Bastian et al.\ 2005).  These are also seen out to $z\sim3$ in
strongly lensed galaxies (Jones et al.\ 2010; Wisnioski et al.\ 2012;
Livermore et al.\ 2012, 2015; Swinbank et al.\ 2012; Adamo et
al.\ 2013; Vanzella et al.\ 2017b; Dessauges-Zavadsky et al.\ 2017).

Star cluster complexes -- often referred to as cluster complexes in
nearby galaxies -- are known to show a range of surface brightnesses
at all redshifts where they are observed, i.e., $z\sim0$-3 (Bastian et
al.\ 2005, 2006; Jones et al.\ 2010; Swinbank et al. 2012; Wisnioski
et al.\ 2012; Rodr{\'{\i}}guez-Zaur{\'{\i}}n et al.\ 2011; Kennicutt
et al.\ 2003).  Star cluster complexes are also described as
star-forming clumps (or giant HII regions) when observed in distant
galaxies.  A simple fit to the mean surface brightness of star cluster
complexes as function of redshift yields the following relation
(Livermore et al.\ 2015):
\begin{equation}
\log \left( \frac{\Sigma_{clump}}{M_{\odot}\, \textrm{yr}^{-1}\, \textrm{kpc}^{-2}} \right) = (3.5\pm0.5)\log(1+z)-(1.7\pm0.2)
\label{eq:sscevol}
\end{equation}
While many other observations of star cluster complexes at
intermediate to high redshifts are also consistent with the above
trend (Franx et al.\ 1997; Swinbank et al.\ 2009; Wuyts et al.\ 2014;
Johnson et al.\ 2017), some star cluster complexes at $z\sim0$ have
been reported to show much higher (by factors of $\sim$100) surface
densities of star formation (Fisher et al.\ 2017).

The implied evolution in the surface brightness of star cluster
complexes is essentially identical to what one would infer from
dimensional arguments.  The sizes of collapsed sources is generally
found to scale as $(1+z)^{-1}$ (e.g., Bouwens et al.\ 2004; Oesch et
al.\ 2010; Ono et al.\ 2013; Holwerda et al.\ 2015; Shibuya et
al.\ 2015) and the evolution in dynamical time goes as $(1+z)^{-1.5}$,
such that $\Sigma_{SFR} \propto r^{-2} t_{dyn} ^{-1}$ $\propto
(1+z)^{3.5}$.  Nevertheless, it should be recognized that the best-fit
evolution in $\Sigma_{clump}$ with redshift likely suffers from
surface brightness selection effects (given the bias towards selection
of the highest surface brightness star cluster complexes at a given
redshift), so the evolution suggested by Eq.~\ref{eq:sscevol} should
only be considered indicative.  The black and magenta lines indicate
the region in parameter space where we would expect star cluster
complexes and star clusters, respectively, to reside (see
Figure~\ref{fig:msre00}: \textit{right panel}).

We include a gray-shaded trapezoid in Figure~\ref{fig:msre00} to show
the region in size-luminosity parameter space that star cluster
complexes in $z\sim0$-3 galaxies have been found to inhabit.  The
light gray region shows an extrapolation of this relation to $z=3$-6.
The solid red circles correspond to the measured sizes and equivalent
$UV$ luminosities of the star cluster complexes identified in the
highly magnified $z=4.92$ galaxy behind MS1358+62 (Franx et al.\ 1997)
by Swinbank et al.\ (2009) and Jones et al.\ (2010), while the solid
cyan circle and green triangle correspond to the sizes and
luminosities of IIZw40 and 30 Doradus, respectively (English \&
Freeman 2003; Vanzi et al.\ 2008).

\begin{figure*}
\epsscale{0.94} \plotone{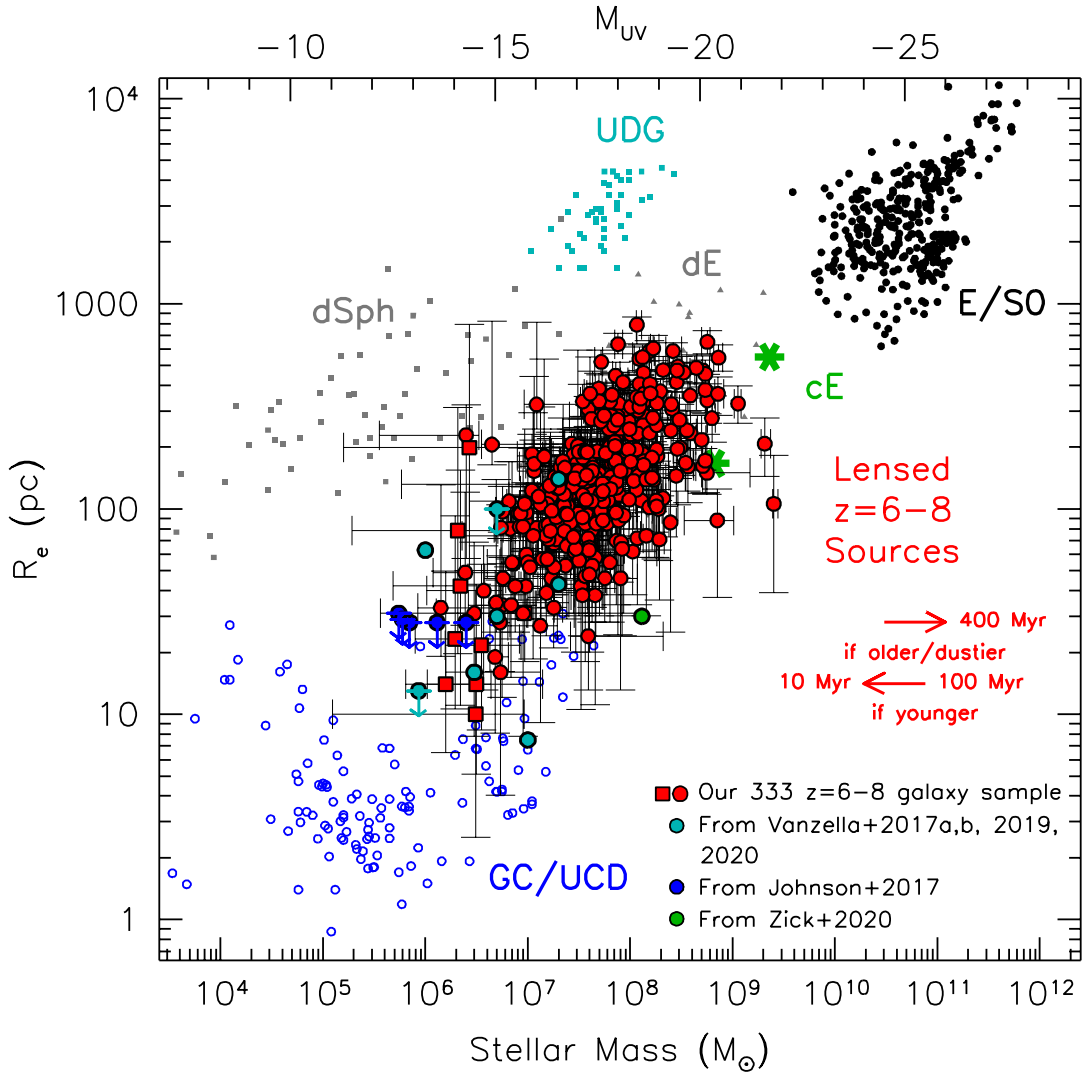}
\caption{Similar to Figure~\ref{fig:msre_sc}, but focusing on
  comparisons with the sizes and masses of various evolved stellar
  systems in the nearby universe, including E/S0 (\textit{black
    circles}), ultra diffuse elliptical galaxies (UDG: \textit{cyan
    squares}), dwarf spheroids (dSphs: \textit{gray squares}), dwarf
  ellipticals (dEs: \textit{gray triangles}), compact ellipticals
  (cEs: \textit{green star-like symbols}), and globular clusters/ultra
  compact dwarfs (GC/UCD: \textit{open blue circles}).  The conversion
  between a given $UV$ luminosity and a stellar mass is made assuming
  a star formation duration of 100 Myr (as found to be appropriate by
  Kikuchihara et al.\ 2020).  The left and right red arrows shows the
  expected change in the inferred masses when changing the duration of
  star formation from 100 Myr to 10 Myr and 400 Myr, respectively.
  The layout of this figure is similar to Figure 8 in Brodie et
  al.\ (2011) and Figures 11 of Norris et al.\ (2014).  See also
  Figure 10 of Bouwens et al.\ (2017c) and Figures 11-13 of
  Kikuchihara et al.\ (2020).  While the smallest lensed $z=6$-8
  sources in the HFF observations appear to have sizes and
  luminosities consistent with that of globular clusters or
  ultra-compact dwarfs, most of the lensed $z=6$-8 galaxies have sizes
  and luminosities that lie in the region between globular clusters
  and that of elliptical galaxies and seem to best match that seen in
  star cluster complexes (Figure~\ref{fig:msre_sc}).  They presumably
  undergo further dynamical evolution and/or accretion before becoming
  the evolved descendants we see today.\label{fig:msre}}
\end{figure*}

Wisnioski et al.\ (2012) find that surface brightness of star cluster
complexes depend on luminosity as $L^{0.26}$, such that the most
luminous star cluster complexes also have the highest surface
brightnesses.

\subsection{Putting Lensed Star-Forming Sources in Context}

Having presented size, luminosity, and mass relations for star
clusters and cluster complexes, we now compare them with our lensed
samples of $z\sim6$-8 sources from Bouwens et al.\ (2021a).  We present
such a comparison in Figure~\ref{fig:msre_sc}, including also the
candidates identified by Vanzella et al.\ (2017a, 2017b, 2019, 2020),
Johnson et al.\ (2017), and Zick et al.\ (2020).  The sizes and
luminosities in Figure~\ref{fig:msre_sc} (and Figure~\ref{fig:msre})
use values for the linear and total magnification factor of 30 and 50,
respectively, for sources where the median magnification factors from
the lensing models exceed 30 and 50, respectively.

The black and magenta lines indicate the region in parameter space
where we would expect star cluster complexes and star clusters,
respectively, to reside (as we showed in Figure~\ref{fig:msre00}).
Most strikingly, the size and luminosities of our sample of 333 lensed
$z\sim6$-8 sources are very similar overall to what is found for star
cluster complexes in $z\sim2$-3 galaxies.

Interestingly, in a few cases, the overall sizes and luminosities of
these sources even extend into the parameter space occupied by super
star clusters.  It is difficult to be sure about a star cluster
identification at the spatial resolutions available with HST (or
ground-based telescopes) even with lensing magnification.
Nevertheless, there are a few sources that have been identified, i.e.,
the core of the Vanzella et al.\ (2017a) D1 source and knot A of the
Sunburst arc (Dahle et al.\ 2016; Vanzella et al.\ 2020), where there
is particularly strong evidence for the sources having very high
magnification factors and very small sizes thanks to the availability
of multiple images of each source behind a given lensing cluster
(Vanzella et al.\ 2019, 2020).  The core of the D1 source has an
inferred size $<$13 pc, while knot A of the Sunburst arc has a likely
size of 5-10 pc.

We will discuss these similarities in more detail in
\S5.

\section{Connection to Sources in the Local Universe\label{sec:local}}

Beyond exploring potential similarities with star clusters and cluster
complexes, it is interesting to ask how the faint lensed sources we
are finding compare with various stellar systems found in the nearby
universe.

For this exercise, we use the compilation that Norris et al.\ (2014)
and M. Norris (2021, private communication) provide of the sizes and
masses for a wide variety of local sources.  This compilation includes
elliptical and S0 galaxies (e.g., Cappellari et al.\ 2011; McDermid et
al.\ 2015), ultra-diffuse ellliptical galaxies (e.g., van Dokkum et
al.\ 2015), dwarf ellipticals and spheroids (e.g., Misgeld et
al.\ 2008), compact ellipticals such as Messier 32 (e.g., Chilingarian
et al.\ 2009), ultra-compact dwarfs (e.g., Evstigneeva et al.\ 2007;
Misgeld et al.\ 2011), and globular clusters (e.g., Hasegan et
al.\ 2005; Firth et al.\ 2007; Mieske et al.\ 2007; Francis et
al.\ 2012).

Figure~\ref{fig:msre} shows the inferred sizes and masses for our
lensed $z\sim6$-8 sample relative to the Norris et al.\ (2014)
compilation.  Other noteworthy especially small candidates from the
literature (e.g., Vanzella et al.\ 2017a, 2017b, 2019, 2020; Johnson
et al.\ 2017; Zick et al.\ 2020) are also presented.  The indicative
masses that we use for our lensed sample are computed assuming a fixed
stellar population duration of 100 Myr in converting from their
measured $UV$ luminosities $M_{UV}$.  Kikuchihara et al.\ (2020) found
that this stellar population age was a good match to the mass-to-light
ratios derived through detailed stellar population modeling of the
full HST+Spitzer/IRAC photometry from lensed $z=6$-8 sources.

Interestingly, some lensed sources in our samples have size and
luminosities in the regime of ultra-compact dwarf galaxies or globular
clusters, with measured sizes $<$40 pc.  Kawamata et al.\ (2015) had
previously reported two sources with such small sizes.  Now, largely
due to the new compilations by Bouwens et al.\ (2017a), Kawamata et
al.\ (2018), and Bouwens et al.\ (2021a) and candidates by Johnson et
al.\ (2017), Vanzella et al.\ (2017a, 2017b, 2019, 2020), we now
recognize that many, if not most, very faint galaxies at $z\sim4$-8
may be small, but we cannot quantify the fraction yet with certainty
(e.g., Figure 5 from Bouwens et al.\ 2020).  Bouwens et al.\ (2021a)
provides an extensive discussion of the evidence for why many faint
$z\sim6$-8 galaxies seem to be very small, even if the individual
measurements are not yet definitive.

\begin{figure}
\epsscale{1.18} \plotone{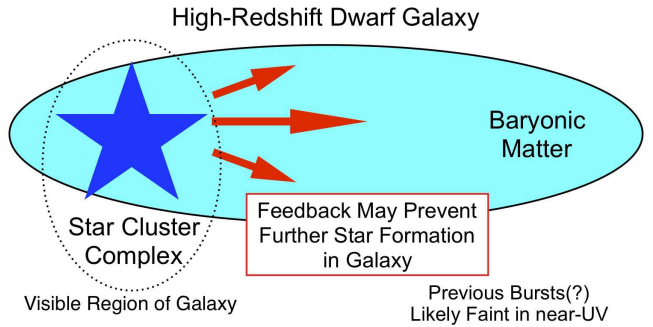}
\caption{Cartoon schematic showing the formation of a single star
  cluster complex within a high-redshift dwarf galaxy.  The star
  cluster complex forms out of overdense baryonic material.  As
  feedback from the star cluster complex could temporarily inhibit
  star formation in other regions of the galaxy (e.g., Bastian 2008;
  Ploeckinger et al.\ 2019), galaxies may be much smaller in terms of
  their readily-visible spatial extent than they actually are.  The
  spatial size of distant star-forming galaxies would be made to look
  smaller than they are due to dominant impact of the youngest star
  cluster complexes on the $UV$ morphologies of dwarf galaxies (e.g.,
  Overzier et al.\ 2008; Ma et al.\ 2018), as occurs e.g. in nearby
  tadpole galaxies (e.g. Kiso 5639: Elmegreen et al.\ 2016) or blue
  compact dwarf galaxies (Elmegreen et al.\ 2012b; Papaderos et
  al.\ 2008).  In particular, Figures 4 and 5 of Elmegreen et
  al.\ (2016) would be strikingly similar to the compact objects seen
  at high redshift.  The lensed $z\sim5$ galaxy MS1358+62 (Franx et
  al.\ 1997), with a dominant star cluster complex $<$200 pc in size
  (see Swinbank et al.\ 2009; Zitrin et al.\ 2011), is another
  dramatic example.\label{fig:cartoon}}
\end{figure}

\section{Discussion}

\subsection{Just a Single Prominent Star Cluster or Star Cluster Complex Per Faint Galaxy?\label{sec:starcluster}}

Interestingly most of the lensed $z\sim6$-8 sources behind the HFF
clusters show very similar sizes and luminosities as star cluster
complexes identified in $z\sim2$-3 galaxies
(Figure~\ref{fig:msre_sc}); in particular, they are significantly
smaller in size than the extrapolated size-luminosity relation for
$z=6$-8 galaxies from blank field studies where radius scales as
$L^{1/4}$ or $L^{1/3}$ (e.g., de Jong \& Lacey 2000; van der Wel et
al.\ 2014).

As has been speculated in the past (e.g., Bouwens et al.\ 2017a;
Vanzella et al.\ 2017a), one might ask if we are simply seeing
individual star cluster complexes in $z\sim6$-8 galaxies and not the
galaxies as a whole.  Only a fraction of the baryonic material
associated with dwarf galaxies, could be lighting up in the rest-$UV$
in the observations, i.e., we might be seeing a single dominant star
cluster complex (e.g., Figure~\ref{fig:cartoon}).  In the Overzier et
al.\ (2008) study of local Lyman-Break Analogue galaxies, it was noted
how our entire rest-$UV$ view of these galaxies could be dominated by
a few cluster complexes.  Overzier et al.\ (2008) speculated about how
fainter Lyman-break analogues might look and whether they might be
dominated by a single super star cluster or cluster complex (see their
\S5.2.2).

More recent examples of galaxies where the rest-$UV$ view of galaxies
are dominated by a small region of the galaxy include Kiso 5639
(Elmegreen et al. 2016).  At lower luminosities closer to the $z=6$-8
galaxies we are considering, possible $z\sim0$ analogues could include
the recent extremely metal poor, high sSFR galaxies shown in Figure 1
of Isobe et al.\ (2020).  There metal poor galaxies appear to be
significantly less spatially extended than the seemingly associated
galaxies they seem to be a part.  While no rest-$UV$ observations are
presented by Isobe et al.\ (2020) to evaluate their spatial extent in
the rest-$UV$, the very strong line emission from these galaxies
suggests they may dominate the morphology of these galaxes at such
wavelengths, making such sources very compact in the rest-$UV$, with
sizes predominantly in the range 40 to 500 pc.  Consistent with the
suggested clumpy star-cluster complex scenario, observations show an
increasing fraction of light in compact star-forming regions, from low
redshift to high redshift (Ribeiro et al.\ 2017).

How likely is it for lower-mass galaxies in the $z=6$-8 universe to
host just a single dominant star cluster complex?  There are several
recent observational and theoretical results we can consider to try to
address this question.

From a theoretical perspective, addressing this question properly
would require high resolution hydrodynamical simulations (e.g., Ma et
al.\ 2018, 2020; Ploeckinger et al.\ 2019).  Nevertheless, one
plausible scenario is to imagine the collapse of an overdensity
resulting in the formation of a star cluster complex and feedback from
that star cluster complex preventing star formation from occurring at
any other position in a dwarf galaxy (Figure~\ref{fig:cartoon}).
While SNe winds might be one possible mechanism (i.e., 4 Myr feedback
time scale for wind speeds of 50$\,$km/s and source sizes of $\sim$200
pc: see Bastian 2008), another mechanism for inhibiting star formation
in other parts of lower mass galaxies would be through radiative
feedback.  Ploeckinger et al.\ (2019) show that this mechanism could
well be more effective at inhibiting star formation in other regions
of lower-mass galaxies.

Zick et al.\ (2018) looked in detail at the probable appearance of
forming globular clusters in the Fornax dwarf galaxy using information
in the color magnitude diagrams of field stars in Fornax and stars in
its 5 globular clusters.  Zick et al.\ (2018) demonstrated that the
forming globular clusters could have been 10-100 times brighter than
the Fornax dwarf galaxy as a whole during its formation epoch and
$>$10$^4$ brighter in terms of their surface brightness, implying that
the only readily detectable regions of dwarf galaxies in deep
extragalactic views are the forming globular clusters themselves.

New simulation results from Pfeffer et al.\ (2019) using the E-MOSAICS
(MOdeling Star cluster population Assembly in Cosmological Simulations
with EAGLE: Pfeffer et al.\ 2018; Kruijssen et al.\ 2019) project
offer a somewhat different perspective on the detectability of forming
globular clusters.  Given the sophistication of the E-MOSAICS project
and the care taken to incorporate star cluster formation, evolution,
and disruption into the EAGLE galaxy formation simulations (Schaye et
al.\ 2015), results from the project should give us a fairly realistic
gauge of what we are observing in current data sets.  Most notably,
Pfeffer et al.\ (2019) find that star clusters form in those regions
of galaxies where substantial clustered and unclustered star formation
is taking place.  This would suggest that compact star-forming regions
in the distant universe most likely correspond to cluster complexes
and not proto-globular clusters forming in isolation.  It is unclear
at present how to reconcile these results with the inferences made by
Zick et al.\ (2018) for globular clusters forming in the Fornax dwarf
galaxy.

While hypothesizing that a single star-cluster complex may dominate
the rest-$UV$ view of many lensed $z=6$-8 galaxies, we suspect that
the underlying host galaxies are likely larger in terms of their
physical extent (cf., the $z=4.92$ Franx et al.\ 1997 MS1358+62
example).  It would be easy for the lower surface brightness regions
to be missed due to cosmic surface brightness dimming (e.g., see Ma et
al.\ 2018).

\subsection{Limits on the Volume Densities of Forming Globular Clusters in the $z\sim6$ Universe\label{sec:gclf}}

We can also use the size measurements from Bouwens et al.\ (2021a) to
set constraints on the luminosity function of proto-globular clusters
in the early universe.  Clearly, the large ages of stars in most
globular clusters together with the high gas densities appropriate for
globular cluster formation (Goddard et al.\ 2010; Adamo et al.\ 2011;
Silva-Villa \& Larsen 2011) -- as well as the high prevalence of
globular clusters even in lower-mass galaxy halos (Spitler \& Forbes
2009; Harris et al.\ 2013) -- strongly suggest a $z\gtrsim 1.5$
formation era.  Having observational constraints on the formation of
these sources in the early universe is both valuable and interesting.

Given the proximity in time of powerful facilities like the James Webb
Space Telescope JWST, there are now numerous predictions for the
number of such clusters which might be found in a typical search field
with the JWST (Renzini 2017; Boylan-Kolchin 2017, 2018; Elmegreen et
al.\ 2012a) as well as candidate proto-globular clusters identified in
separate studies (Vanzella et al.\ 2017a).

\begin{deluxetable}{cc}
\tablecolumns{2}
\tabletypesize{\footnotesize}
\tablecaption{Volume Density Constraints on proto-globular clusters forming at $z\sim6$\label{tab:gclf}}
\tablehead{\colhead{$M_{UV}$} & \colhead{$\phi$ (Mpc$^{-3}$)}}
\startdata
\multicolumn{2}{c}{This Work\tablenotemark{a}}\\
$-$17.50 & $<$0.0001\tablenotemark{b}\\
$-$16.50 & $<$0.0007\tablenotemark{b}\\
$-$15.50 & $<$0.004\tablenotemark{b}\\
$-$14.50 & $<$0.047\tablenotemark{b}\\
$-$13.50 & $<$0.14\tablenotemark{b}\\
\\
\multicolumn{2}{c}{Estimated From Vanzella et al.\ (2017a)}\\
$-$15.50 & $<$0.0106\tablenotemark{b,c}\\
\\
\multicolumn{2}{c}{Volume Densities Probed with the HFF program\tablenotemark{d}}\\
$-$17.50 & 0.00014\\
$-$16.50 & 0.00041\\
$-$15.50 & 0.0020\\
$-$14.50 & 0.011\\
$-$13.50 & 0.062\\
$-$12.50 & 0.9
\enddata
\tablenotetext{a}{These upper limits are derived using Bouwens et
  al.\ (2017b) $z\sim6$ LF results and the fraction of faint $z\sim6$
  galaxies whose measured size is 40 pc or smaller.  These upper
  limits are based on median linear magnification factors from the
  parametric models.  If we only include those 21 sources (out of 23)
  which satisfy our $<$40 pc-criterion when utilizing a maximum linear
  magnification factor of 30, the upper limits we derive would be 10\%
  more stringent than presented here.}
\tablenotetext{b}{$1\sigma$ upper limits}
\tablenotetext{c}{We use the same criteria in establishing the upper limits on the proto-globular cluster volume densities as we use for the Bouwens et al.\ (2021a) observational results}
\tablenotetext{d}{Gray region in Figure~\ref{fig:gclf}}
\end{deluxetable}

\begin{figure}
\epsscale{1.1}
\plotone{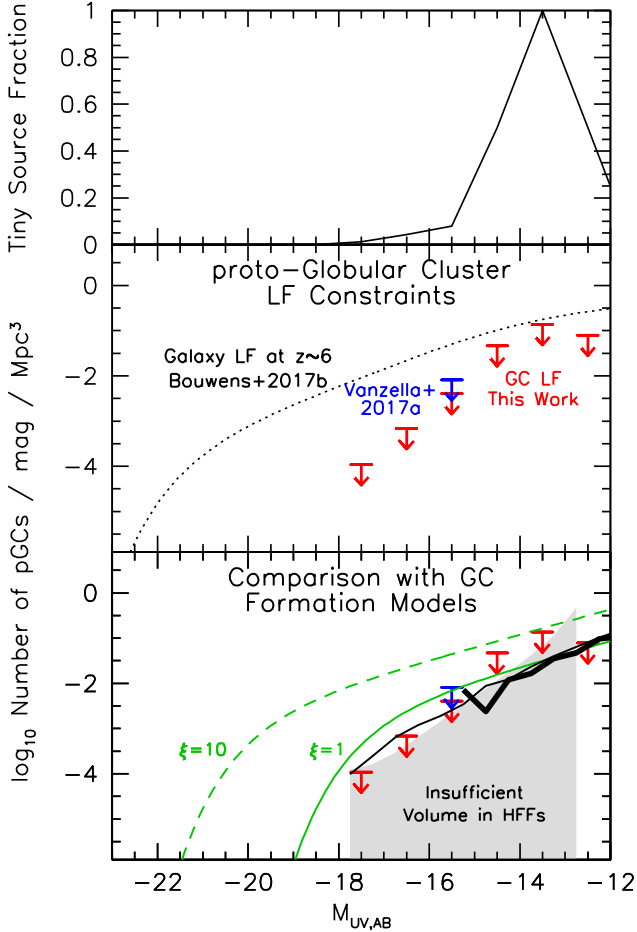}
\caption{(\textit{upper}) Fraction of the lensed $z\sim6$ sources in
  the HFF observations which are measured by Bouwens et al.\ (2021a) to
  have half-light radii $\leq$40 pc.  (\textit{middle}) Constraints on
  the volume density of forming proto-globular clusters at $z\sim6$
  using searches for small sources behind the HFF clusters.  The
  plotted red upper limits combine the $UV$ LF constraints we obtained
  from all six HFF clusters with the fraction of sources for which the
  available constraints suggest sizes $\leq$40 pc.  The blue upper
  limit gives the volume density constraints we infer here for the
  proto-globular cluster reported by Vanzella et al.\ (2017a).  The
  dotted black line shows a recent determination of the $z\sim6$ LF
  for galaxies from the Hubble Frontier Fields from Bouwens et
  al.\ (2017b).  This LF was used to help derive the observational
  constraints (together with the results in the top panel).
  (\textit{lower}) Comparison of the observational constraints on the
  proto globular cluster volume densities with the predicted LF of
  proto-globular cluster candidates from J. Pfeffer et al.\ (2021, in
  prep: \textit{thin black line}), Pfeffer et al.\ (2019:
  \textit{thick black line}), and Boylan-Kolchin (2017, 2018) assuming
  the mass ratio $\xi = 1$ (\textit{green solid line}: see \S\ref{sec:gclf}) and
  $\xi = 10$ (\textit{green dashed line}), where $\xi$ is the ratio of
  stellar mass in the initial globular cluster at its birth and that
  present at $z=0$.  The shaded gray region shows the volume densities
  and luminosities where the HFF program does not provide us with
  sufficient volume to probe; its upper envelope is equal to the
  reciprocal of the total volume computed to be available over the
  first six HFF clusters in the Bouwens et al.\ (2021, in prep)
  analysis.\label{fig:gclf}}
\end{figure}

To provide constraints on the volume density of forming globular
clusters in the early Universe, we explicitly consider the size
constraints we have available for the full sample of $z\sim6$ sources
over all six HFF clusters from Bouwens et al.\ (2021a) vs. that
expected for star clusters (Figure~\ref{fig:msre_sc}).  If the size
measurements for a source yield a half-light radius measurement of
$<$40 pc, we consider it as a possible globular cluster candidate.  In
this discussion we use the full sample of sources with sizes $<$40 pc
from Bouwens et al.\ (2021a), but also show the (small) impact of
limiting ourselves to linear and total magnifications of 30 and 50,
respectively.  This target size range factors in the $\sim$10-20-pc
maximum observed sizes of globular clusters
(\S\ref{sec:intro_starcluster}), with a few notable exceptions (Pal 5
and Pal 14: Baumgardt \& Hilker 2018; Kikuchihara et al.\ 2020) and a
factor-of-$\sim$2 evolution in the size of the globular clusters
themselves that can occur following their initial formation time.
Such size evolution could occur due to two-body relaxation, mass loss
by stellar evolution, and galaxy tides (Hills 1980; Goodman 1984;
Baumgardt \& Makino 2003; Shin et al.\ 2013).  A $<$40-pc size limit
also allows us to be as inclusive as possible in selecting possible
globular cluster candidates to ensure that the upper limit we set on
the volume density of forming globular cluster candidates is robust.
This is a careful, even conservative approach, that is particularly
appropriate with our sample given the significant systematic and
random uncertainties in our direct size measurements, as we discussed
extensively in Bouwens et al. (2021a).  Using this definition, one of
the $z\sim6$ sources with size measurements from Kawamata et
al.\ (2015), i.e., HFF1C-i10, would already qualify as a
proto-globular cluster candidate.

In the upper panel of Figure~\ref{fig:gclf}, we present the fraction
of $z\sim6$ sources which could correspond to forming globular
clusters vs. $UV$ luminosity (i.e., those with $r_e \leq 40$ pc: see
Table~\ref{tab:tiny}).  We then show in the middle panel of
Figure~\ref{fig:gclf}, the implied upper limits we obtain on the
volume density of proto-globular clusters forming in the distant
universe.  Given the challenges in being sure that any given system
corresponds to a proto-globular cluster (or the challenges in being
sure that specific lensed sources are in fact small: see \S4.4-4.5
from Bouwens et al.\ 2020), we include our constraints as upper
limits.  The upper limits and fractions we present in
Figure~\ref{fig:gclf} are based on median linear magnification factors
from the parametric models.  If we only include those 21 sources (out
of 23) which satisfy our $<$40 pc-criterion when utilizing a maximum
linear magnification factor of 30, the upper limits we derive would be
10\% more stringent than what present in Table~\ref{tab:tiny} and
Figure~\ref{fig:gclf}.

In the same panel, we also include the one proto-globular cluster
candidate GC1 identified by Vanzella et al.\ (2017a) at $z\sim6$ again
as an upper limit on the volume density (see also Vanzella et
al.\ 2019).  For consistency with the candidates included from our own
study, we only include this candidate since its inferred size of
16$\pm$7 pc satisifes our criterion of a source size $<$40 pc.  In
Table~\ref{tab:gclf}, we provide the estimated upper limits on the
volume density of sources for comparison.

An important uncertainty in attempting to quantify the volume density
of globular clusters forming in the distant universe regards the
possibility that globular clusters may not be identifiable separately
from the host galaxy of which they are a part.  As we discusssed in
\S\ref{sec:starcluster}, Zick et al.\ (2018) made use of the available
resolved stellar population information on the Fornax dwarf galaxy to
demonstrate that globular clusters forming in that system could be
clearly identifiable as separate systems during their formation phase.
The E-MOSAICS simulation results presented by Pfeffer et al.\ (2019)
suggest that such identifications could be much more challenging,
however, due to star clusters typically forming in significant
associations.  While such could make the identification of single star
clusters more challenging, we will need to postpone a detailed
consideration of these issues to a future analysis where we can
simulate these issues in detail in the context of a full globular
cluster+galaxy formation simulation like E-MOSAICS.

For comparison with our proto-globular cluster constraints, we also
include on the lower panel of Figure~\ref{fig:gclf} several
predictions for the number density of proto-globular cluster and
evolved globular cluster systems as a function of $UV$ luminosity
present in the $z\geq 6$ universe using several different
proto-globular cluster formation models.  The first predictions are
from the E-MOSAICS project (Pfeffer et al.\ 2018; Kruijssen et
al.\ 2019), which explicitly adds star cluster formation to the EAGLE
galaxy formation simulations, while matching of the observed scaling
relations and prevalence of young and evolved star clusters.  Using
results from E-MOSAICS executed on a (34.4 comoving Mpc)$^3$ volume,
we show predictions for the $UV$ luminosity function of the forming
globular clusters at $z\sim6$ (\textit{thin black line}: J. Pfeffer et
al.\ 2020).  Also shown are the predicted from Pfeffer et al.\ (2019)
zoom-in simulation results (\textit{thick black line}) on 25
Milky-Way-like galaxies normalized to match the new J. Pfeffer (2021,
in prep) results at the low luminosity end.

Additionally, we show the predicted globular cluster $UV$ LF from the
Boylan-Kolchin (2017, 2018) model.  The Boylan-Kolchin (2018) approach
has particular value due to its relative simplicity and the convenient
Schechter function approximation provided for the proto-globular
cluster LF.  Shown are the predicted globular cluster LF for two
different values of the ratio of stellar mass in the initial globular
cluster at its birth and that present at $z=0$, i.e., $<m_{GC}
(\textrm{birth})>/<m_{GC} (z=0)>$.  Following Boylan-Kolchin (2017,
2018), we use the symbol $\xi$ to describe this ratio and present both
the $\xi=1$ case (where $\phi^*=4 \times 10^{-3}$ Mpc$^{-3}$,
$M^*=-16.9$, $\alpha=-1.7$) and $\xi=10$ case (where $\phi^*=4 \times
10^{-3}$ Mpc$^{-3}$, $M^*=-19.4$, $\alpha=-1.7$) in
Figure~\ref{fig:gclf}.  The $\xi=10$ case involves substantial mass
loss after the initial globular burst would favor very bright
proto-globular clusters in the early universe.  Such a scenario is
motivated e.g. by Schaerer \& Charbonnel (2011) based on chemical
complexity of the enrichment in globular clusters (see also D'Ercole
et al.\ 2008; Renzini et al.\ 2015).  E-MOSAICS features very little
mass loss from forming globular clusters and therefore the effective
mass-loss factor $\xi$ is closest to 1.  The predicted $UV$ LFs of
other models in the literature, e.g., Pozzetti et al.\ (2019), lie
between the $\xi=1$ and $\xi=10$ results from Boylan-Kolchin (2017,
2018).

For context, we also show on the lower panel of Figure~\ref{fig:gclf}
the volume densities to which we would be able to search for
proto-globular clusters of specific luminosities with the full HFF
program.  The volume densities are computed as in Bouwens et
al.\ (2017b) and hence would be for a probe of proto-globular clusters
at $z\sim6$.  They are based on the results over all six HFF clusters
(Bouwens et al.\ 2021, in prep).  These volume densities are also
compiled in Table~\ref{tab:gclf} for convenience.  The search volume
available for proto-globular clusters in the $z=6$-10 universe is
$\sim$3$\times$ larger than in the $z\sim6$ universe (i.e.,
$z\sim5.5$-6.3 using the Bouwens et al.\ 2017b criteria).

Remarkably, the predictions from E-MOSAICS (Pfeffer et al.\ 2019;
J. Pfeffer et al.\ 2021, in prep) lie quite close to the upper limits
we can set on the basis of existing HFF search results at $-$15 mag
and brightward.  E-MOSAICS predicts $\sim$1.5 proto-globular cluster
candidates at both $-$17.5 and $-$16.5 mag in the total magnified
volume behind the HFF clusters, consistent with the number of
candidates found here from the Bouwens et al.\ (2021a) compilation.
The $\xi=1$ model from Boylan-Kolchin (2017, 2018) predicts a
proto-globular cluster LF that is $\sim$0.4-0.5 dex higher than the
E-MOSAICS model.  This is clearly in excess of our HFF constraints
brightward of $-$16 mag, predicting $\sim$3 and $\sim$10 sources at
$-17.5$ mag and $-16.5$ mag, respectively, within the volume of the
HFF program.  As mentioned above, the Pozzetti et al (2019) model
predictions lie between the $\xi = 1$ and the $\xi = 10$ lines.  As
such the Pozzetti et al. (2019) model predicts an even larger number
of proto-globular cluster candidates ($\approx$2-3$\times$ larger) to
be present in the HFF volume (not shown in Figure~\ref{fig:gclf}).

At slightly fainter luminosities, i.e., $-$15.5 mag, our observational
results (\textit{red downward arrows}) are also more consistent with
the minimal $\xi=1$ scenario sketched out by Boylan-Kolchin (2017:
\textit{green solid line}).  If the Boylan-Kolchin (2018) $\xi=1$
scenario is correct, three $-15.5$-mag sources identified with the HFF
program are expected to correspond to proto-globular clusters in
formation.  For the E-MOSAICS model, one $-15.5$-mag source would
correspond to a proto-globular cluster.  While still small numbers,
these constitute $\sim$50\%-100\% of the tiny star-forming sources
that we have identified at those low luminosities.  These results
indicate that we observe plausible consistency between the current
globular cluster formation models and what we derive from our size and
LF results.  Faintward of $-$14 mag, the predicted proto-globular
prevalence from the E-MOSAICS and Boylan-Kolchin (2017, 2018) models
appear to be too low to be well probed by the HFF program.

Interestingly, and rather definitively, the $\xi=10$ model of
Boylan-Kolchin (2017) exceeds the upper limits we can set from the HFF
observations at all luminosities.  As such, we can likely already rule
the $\xi=10$ model out.  This adds to other independent
evidence against such models (Bastian \& Lardo 2015; Kruijssen 2015;
Webb \& Leigh 2015; Martocchia et al.\ 2017; Elmegreen 2017).

The present comparisons suggest that observers may be on the brink of
exploring the formation of globular clusters in the distant universe
with current and especially using future observations with JWST (see
also discussion in Renzini 2017, Vanzella et al.\ 2019, and Pozzetti
et al.\ 2019).  With JWST, not only will we be able to search for
proto-globular cluster candidates much more efficiently, but we will
be able to characterize each candidate system in detail using high S/N
spectrocopy, probing the velocity dispersion (and thus allowing for a
measurement of the dynamical mass) as well as the chemical maturity of
such systems.

\subsection{Challenges Related to the Size Measurements We Utilize\label{sec:small}}

The comparisons we present here of source sizes at $z\sim 6$-8 and
those of local star forming regions exemplifies the power of using
lensing clusters.  However, as is discussed in \S4.5 of Bouwens et
al. (2021a), uncertainties in the magnification factors can have a
significant impact on the measured sizes, resulting in sources whose
sizes are both larger and smaller than reality.  The tests discussed
in Bouwens et al. (2021a), which build on the work of Bouwens et
al.\ (2017a) and (2017b), suggest that recovery of the true sizes and
luminosities of the original lensed galaxies is nevertheless possible
up to linear magnification factors of $\sim$30 (for sizes), and total
magnification factors of $\sim$50 (for luminosities), but become less
unreliable at higher magnification factors.  The collective impact of
uncertainties in the lensing models on the derived size distributions
is illustrated well with Figure 5 of Bouwens et al. (2021a).

Nonetheless, there is an abundance of evidence that many faint sources
are small, and that the broad distribution of sources we show here in
Figures~\ref{fig:msre_sc} and \ref{fig:msre} are likely representative
of the real distribution of sources.  One especially strong piece of
evidence is from the observed surface density of faint $z\sim6$-8
galaxies in the highest magnification regions behind lensing clusters
and the fact that this surface density appears to show no dependence
on the shear factor (Bouwens et al.\ 2017a).  This could only be the
case if faint galaxies had intrinsically small sizes, i.e., $<$165 pc
(86\% confidence: Bouwens et al.\ 2017a).

Second, a number of sources with especially small size measurements
benefit from the existence of other multiple images of the same
source.  The existence of such images allow us to set tighter
constraints on the magnification of sources at very high magnification
factors.  Examples of such sources include both the central
proto-globular candidate in D1 (Vanzella et al.\ 2017, 2019) and the
candidate YMC in knot A of the Sunburst arc (Vanzella et al.\ 2020)
and give us confidence that many more such small sources exist amongst
the lensed population of $z=6$-8 galaxies behind the HFF clusters.

Third, a high prevalence of faint $z=6$-8 galaxies is found in high
magnification regions behind the HFF clusters (Bouwens et
al.\ 2017a,b; Ishigaki et al.\ 2018; Bouwens et al.\ 2020).  This
could only be the case if sources were significantly smaller than
found from a simple extrapolation of the Shibuya et al.\ (2015)
relation (unless the LF showed a concave-upwards form, which seems
inconsistent with galaxy formation models).  The discussion in
\S5.3 from Bouwens et al.\ (2021a), as well as their Figure 10,
illustrates this well.

Finally, there are also the size-luminosity results at $z\sim4$
(Bouwens et al.\ 2020) which are less impacted by incompleteness
systematics and provide clearer evidence for lower luminosity sources
having very small sizes.  While there are a number of uncertainties in
the size-luminosity relation we derive from direct measurements, the
above results collectively provide strong evidence that (1) galaxies
fainter than about $-$17 $M_{UV}$ have very small sizes, and (2) do so
with a slope that appears to be steeper than for luminous galaxies.
An important caveat here is that our size measurements are in the
rest-$UV$ and thus may not reflect the true spatial extent of a
galaxy, but merely that of the dominant star forming region, as we
discussed in \S\ref{sec:starcluster} and showed schematically in
Figure~\ref{fig:cartoon}.

\section{Summary}

We have made use of a significant sample of 333 lensed star-forming
galaxies behind the HFF clusters and compared their inferred sizes and
luminosities to various classes of sources in the local universe to
help put these high-redshift sources in context.  Both the
identification of these $z\sim6$-8 sources and their size measurements
are described in a companion paper (Bouwens et al.\ 2021a).

The results presented here broadly assume that that the distribution
of sizes and luminosities inferred by Bouwens et al.\ (2021a) are
representative of reality.  Indeed, there is a wide variety of
indirect evidence supporting the general conclusion from Bouwens et
al.\ (2021a) that most lower luminosity, $z\sim6$-8 sources are small,
as discussed here in \S5.3.  Nevertheless, it is worth being aware of
the significant impact that issues like incompleteness and
uncertainties in the lensing models can have on the measured sizes and
luminosities, as e.g. discussed by Bouwens et al.\ (2021a).  The impact
of these uncertainties is especially large for the lowest luminosity
sources, so some caution is required in interpreting the present
results.

Our analysis shows that $z=6$-8 sources from the Bouwens et
al.\ (2021a) samples have measured sizes and luminosities very similar
to that derived for star cluster complexes identified in galaxies at
$z=0$-3 (Jones et al.\ 2010; Livermore et al.\ 2012, 2015; Wisnioski
et al.\ 2012; Swinbank et al.\ 2012; Johnson et al.\ 2017).  In fact,
the typical $-$15 mag galaxy in our samples has a smaller half-light
radius than 30 Doradus, which has a measured half-light radius of
$\sim$100 pc.  This could be interpreted to suggest that lower
luminosity galaxies in the early universe may often contain a single
prominent star cluster complexes which dominates the observed UV
morphology (Figure~\ref{fig:cartoon}).

We also placed the measured size and luminosities of lensed $z=6$-8
galaxies in our samples with the sizes and masses of stellar systems
in the nearby universe (\S4).  Most of the sources have
inferred masses and luminosities that place them in the region of
parameter space where star cluster complexes lie
(Figure~\ref{fig:msre_sc}), which occurs midway between ultra-compact
dwarfs and elliptical galaxies (Figure~\ref{fig:msre}).  This suggests
that many low-luminosity galaxies may be dominated by a single star
cluster complex in terms of their observed morphologies
(Figure~\ref{fig:cartoon}).

Nevertheless, we remark that for a small minority of sources in our
sample, their properties are consistent with potentially corresponding
to super star clusters and -- as such -- they could correspond to
proto-globular clusters (Figure~\ref{fig:msre_sc}: \S3).  There are 23
sources in the Bouwens et al.\ (2021a) selection with median size
measurements equal to 40 pc or smaller.  There are 3 sources from the
Bouwens et al.\ (2021a) selection that are likely to have sizes as
small as 10-20 pc.

We combine current constraints on the fraction of especially small
sources behind the HFF clusters with new state-of-the-art constraints
on the $UV$ LF of sources at $z\sim6$ from the HFF clusters (Bouwens
et al.\ 2017b) to derive constraints on the proto-globular cluster LF
at high redshift (\S\ref{sec:gclf}).  Comparing this LF with
predictions from the recent E-MOSAICS project (Pfeffer et al.\ 2019;
J. Pfeffer et al.\ 2021, in prep), models from Boylan-Kolchin (2017,
2018), and Pozzetti et al.\ (2019), we find that with current
observations from the HFF clusters we are probably very close to
identifying bona-fide globular clusters in formation in the early
universe (if some such sources have not been identified already with
our probe or that of Vanzella et al.\ 2019, 2020).

For example, the E-MOSAICS project predicts that $\sim1.5$
proto-globular clusters in formation should be visible in the HFF
observations at $-$17.5 mag and $-$16.5 mag, respectively, while the
$\xi=1$ model of Boylan-Kolchin (2017, 2018) predicts $\sim$3,
$\sim$10, and $\sim$3 forming proto-globular clusters at $-$17.5 mag,
$-$16.5 mag, and $-$15.5 mag.  Tantalizingly enough, our analysis of
the Bouwens et al.\ (2021a) observational results shows some candidates
with plausible proto-globular cluster sizes at $<$$-$15 mag (see
lowest panel of Figure~\ref{fig:gclf}) in the same luminosity range
where such sources are predicted.

Despite plausible consistency between our search results and the
predictions of e.g. the E-MOSAICS project, the present results already
place strong constraints on more extreme globular cluster formation
scenarios, e.g., those involving substantial ($\xi=10$) mass loss
after the initial formation burst.  Our results appear to rule out
those scenarios entirely.  While this inference is contingent on
forming proto-globular clusters having sufficiently high surface
brightness to be identified as separate from the host galaxy (see also
Bastian \& Lardo 2015; Kruijssen 2015; Webb \& Leigh 2015; Martocchia
et al.\ 2017 for further evidence), such issues are unlikely to cause
all such forming globular clusters to be hidden.

As in our own previous work or other work (i.e., Bouwens et
al.\ 2017a, 2021a; Vanzella et al.\ 2017a, 2017b, 2019; Johnson et
al.\ 2017), one should treat the present results with some caution
(\S\ref{sec:small}) as a result of the impact of uncertainties in the
lensing models.  The present results require that parametric lensing
models be predictive to magnification factors to $\gtrsim$30.
Fortunately, this condition is likely satisfied given the end-to-end
simulation+recovery results obtained by Meneghetti et al.\ (2017) and
own tests comparing the HFF models.

In the future, we aim to gain more insight into the nature of small
star-forming galaxies in the distant universe through more extensive
spectroscopic observations, larger samples, and NIRSPEC/IFU
observations with JWST.

\acknowledgements

We acknowledge stimulating discussions with Angela Adamo, Nate
Bastian, Mike Boylan-Kolchin, James Bullock, Rob Crain, Bruce
Elmegreen, Phil Hopkins, Xaiocheng Ma, Mike Norman, Elliot Quataert,
Alvio Renzini, Britton Smith, Eros Vanzella, Dan Weisz, Shelley
Wright, and Tom Zick.  Nate Bastian and Angela Adamo provided us with
extremely valuable feedback on the scientific content and language
used in an earlier version of this paper, especially with regard to
star clusters and star cluster complexes.  We are grateful to Mark
Norris for sending us a compilation of the sizes, luminosities, and
masses of many evolved stellar systems in the nearby universe.  We
feel very appreciative to Joel Pfeffer for sending us pre-submission
results on the $UV$ LF for proto-globular clusters from the E-MOSAICS
project and providing us with pre-submission feedback on this
manuscript.  We acknowledge the support of NASA grants HST-AR-13252,
HST-GO-13872, HST-GO-13792, and NWO grants 600.065.140.11N211 (vrij
competitie) and TOP grant TOP1.16.057.  This research is based on
observations made with the NASA/ESA Hubble Space Telescope obtained
from the Space Telescope Science Institute, which is operated by the
Association of Universities for Research in Astronomy, Inc., under
NASA contract NAS 5–26555.

\end{document}